\documentclass[11pt]{article}
  
\usepackage[french]{babel}
\usepackage[applemac]{inputenc}

\usepackage{diagrams}
\usepackage{slashed}

\usepackage{graphicx}
\usepackage{amssymb}
\usepackage{epstopdf}
\usepackage{a4wide}
\usepackage{amsmath}
\usepackage{color} 
\usepackage{epic} 
\usepackage{here}

\newcommand{\nco}{\newcommand}
\newfont{\msbm}{msbm10 at 12pt}
\nco{\CC}{\mbox{\msbm C}} 
\nco{\ZZ}{\mbox{\msbm Z}}
\nco{\NN}{\mbox{\msbm N}}
\nco{\HH}{\mbox{\msbm H}}
\nco{\II}{\mbox{\msbm I}} 
\nco{\RR}{\mbox{\msbm R}}
\nco{\munite}{\ensuremath{\,\,\mathrm{l}\!\!\!1}} 
\nco{\dps}{\displaystyle}
\nco{\ov}{\overline}
\nco{\ud}{\underline}

\def\ie{{\rm i.e.,\/}\ }
\def\etc{{\rm etc.\/}\ }

\def\vr#1{\overrightarrow{#1\kern 1pt}\kern-1pt}
\def\vl#1{\overleftarrow{#1\kern 1pt}\kern-1pt}

\title{Clifford algebras, spinors and fundamental interactions :  \\
Twenty Years After \vspace{1.0cm} }

\author{ {\bf R. Coquereaux}
\\
{Centre de Physique Th\'eorique}${}^{}$\thanks{
Unit\'e Mixte de Recherche (UMR 6207) du CNRS
et des Universit\'es Aix-Marseille I, Aix-Marseille II,
et du Sud Toulon-Var; laboratoire affili\'e \`a la
FRUMAM (FR 2291)}}
\date{}
\begin{document}
\begin{titlepage}

\maketitle

\vspace{1cm}

\begin{abstract}

This is a short review of the algebraic properties of Clifford algebras and spinors. 
Their use in the description of fundamental physics (elementary particles) is also summarized.

\smallskip 
\centerline{Lecture given at the ICCA7 conference, Toulouse (23/05/2005).}

\end{abstract}

\vfill


\vspace{1.0cm}

\vspace*{0.5 cm}

\end{titlepage}

\section{Introduction}

The subject of spinors and Clifford algebras  held a kind of fascination for me, when I first met these concepts, long ago. I even published, at that time \cite{Coquereaux : periodicityClifford}, a short letter,  in a physics journal, giving a short proof of the periodicity theorem for arbitrary signatures (to discover later that several references already existed in the mathematical literature !). I also gave, and wrote -- in Austin --   a series of lectures on these topics (published in \cite{Coquereaux: TexasLecturesClifford}), but all that took place more than 20 years ago, and I did not touch the subject since. My surprize to receive an invitation for this conference was therefore great, and  I want to thank the organizers.

They kindly asked me to give a general lecture that would summarize what every (young) theoretical
physicist should know about the subject of Clifford algebras and spinors. It may be that this lecture will be about  what a particular (and old) theoretical physicist still remembers on the subject !
I am also supposed to present an outline of the ``zoology'' of the Standard Model of particle physics, for the benefit of  non-physicists (actually for mathematicians),  trying to explain in a couple of pages how the previous general concepts are used in the conventional\footnote{\ie the accepted corpus of the theory and not its interesting extensions  that would involve  grand unified theories, Kaluza-Klein theories, supersymmetries, strings, branes and the like. } description of elementary particles.

These pedagogical requirements  explain the structure of the talk, which is as follows : Clifford algebras (on the real or complex fields), spinors, Spin groups, spin bundles and spin connections, spinor fields, fundamental interactions and the standard model of elementary particles.

\section{Clifford algebras}


\paragraph{Definition.}

The real Clifford algebra $C(p,q) \equiv Cliff(p,q)$ is the unital associative algebra, generated over the field $\RR$
by $n=p+q$ symbols $\gamma^\mu$ with the relations 
$(\gamma^\mu)^2=1$ for  $\mu \in \{1,2,\ldots,p\}$,  
$(\gamma^\nu)^2=-1$ for $\nu \in \{p+1,p+2,\ldots,p+q\}$, and
$\gamma^\mu \gamma^\nu + \gamma^\nu \gamma^\mu = 0$ when $\mu \neq 
\nu$. The dimension of the Clifford algebra is therefore $2^n$, so that it is isomorphic, as a vector space (not as an associative algebra), with the exterior algebra.

Let $C_0$ denote the even part of  $C$. Call $Z$ the center of  $C$ and $Z_0$ the center of $C_0$.

One defines the orientation operator  $\epsilon = \gamma^0\gamma^1\ldots \gamma^n$.

If  $\epsilon^2 = 1$ we can build the two projectors $P_L = \frac{1-\epsilon}{2}$ and $P_R = \frac{1-\epsilon}{2}$.

\paragraph{Structure of real Clifford algebras : case $n=p+q$ even.} $\quad$

$Z$ is generated by $1$ but $Z_0$ is generated by $1$ and $\epsilon$.

$C$ is a simple algebra.

The discussion depends on the square of $\epsilon$. 

{\vskip 0.5cm}
\begin{eqnarray*}
& C_0\, \hbox{is simple} & \Leftrightarrow \epsilon^2 = -1 
\Leftrightarrow p-q = 2 \,\hbox{ mod}\, 4 \Leftrightarrow Z_0 \sim 
\CC \\
& C_0 \, \hbox{is not simple} &  \Leftrightarrow \epsilon^2 = +1 
\Leftrightarrow p-q = 0 \,\hbox{ mod}\, 4 \Leftrightarrow Z_0 \sim 
\RR \oplus \RR \\
&{}& {\hskip 1cm} \hbox{ in the later case} \quad  C_0 = P_L C_0 
\oplus P_R C_0
\end{eqnarray*}

Remark : for the Dirac algebra (name given to the Clifford algebras when 
$(p=3, q=1)$ or when  $(p=1, q=3)$ ), $\epsilon^2=-1$. The even subalgebra $C_0$ is
simple. In order to decompose it into two direct summands, one needs to use  the complex number
$i$ so that one can build the projectors
$(1 \pm \gamma_5)/2$, with
$\gamma_5 = i \, \epsilon$. This can only be done if we use complex number coefficients, \ie  if we complexify
the algebra $C$ (see later).

\paragraph{Structure of real Clifford algebras : case $n=p+q$ odd.} $\quad$

$Z$ is generated by $1$ et $\epsilon$ but $Z_0$ is generated only by $1$.

$C_0$  is a simple algebra.

The discussion depends on the square of $\epsilon$. 

{\vskip 0.5cm}

\begin{eqnarray*}
& C \, \hbox{is simple} & \Leftrightarrow \epsilon^2 = -1 
\Leftrightarrow p-q = 3 \,\hbox{ mod}\, 4 \Leftrightarrow Z \sim \CC 
\\
& C \, \hbox{is not simple} &  \Leftrightarrow \epsilon^2 = +1 
\Leftrightarrow p-q = 1 \,\hbox{ mod}\, 4 \Leftrightarrow Z \sim \RR 
\oplus \RR \\
&{}&  {\hskip 1cm} \hbox{ in this last case} \quad  C = {1 -  
\epsilon \over 2} C \oplus {1 -  \epsilon \over 2} C
\end{eqnarray*}

Moreover, $C = C_0 \oplus C_0 \epsilon$.

{\bf Remark}

The previous discussion suggests that a classification would depend on the value of $p-q$ mod 4.
A finer analysis will show that it actually depends on $p-q$ mod $8$.

\paragraph{Periodicity theorem.}

There are three ingredients:

\begin{itemize}
\item Isomorphisms for Clifford algebras of low dimension
\item A theorem of dimensional reduction
\item  Standard results for tensor products of the fields $\RR$ (reals), $\CC$ (complex) and $\HH$ (quaternions).
\end{itemize}

\begin{enumerate}
\item  $$C(1,0)=\CC, \quad C(0,1)=\RR \oplus \RR, \quad C(2,0) = M(2,\RR)$$ 
$$C(1,1) = M(2,\RR) \quad  \,\hbox{et} \, \quad C(0,2) = \HH$$
\item If $F = E \oplus E^\perp$ and if the dimension of $E$ is even, then $$C(F,g) = C(E,g_E) \otimes C(E^\perp,\epsilon_E^2 \,   g_{E^\perp})$$ Proof  : define $\Gamma^\mu = \gamma^\mu \otimes 1$,  $\Gamma^\alpha = \epsilon_E \otimes \gamma^\alpha$ and check the defining relations.

\smallskip
Applications :

 if $E$ is a vector space of dimension $2$, its orientation operator is $\epsilon = \gamma^1 \gamma^2$, then 
\begin{itemize}
\item If the signature is $(2,0)$ then $\epsilon^2 = -1$ and 
$C(2,0)\otimes C(p,q) = C(q+2,p)$
\item If the signature is $(0,2)$ then $\epsilon^2 = -1$ and 
$C(0,2)\otimes C(p,q) = C(q,p+2)$
\item If the signature is $(1,1)$ then $\epsilon^2 = +1$ and
$C(1,1)\otimes C(p,q) = C(p+1,p+1)$
\end{itemize}

\item
\begin{eqnarray*}
M(n,\RR) \otimes M(m,\RR) = M(nm,\RR)& \quad &\CC \otimes \CC = \CC 
\oplus \CC \\
M(n,\RR) \otimes \CC = M(n,\CC)& \quad &\HH \otimes \CC = M(2,\CC) \\
M(n,\RR) \otimes \HH = M(n,\HH)& \quad &\HH \otimes \HH = M(4,\RR)
\end{eqnarray*}
Proof :  For $\RR$ and $\CC$, this is obvious. For $H$, write the quaternions as $2\times 2$ matrices 
$\begin{pmatrix}  a-id & ib + c \\ ib - c & a+ id\end{pmatrix} $
\end{enumerate}

{\bf The periodicity theorem.} The previous results imply immediately the following :
\begin{eqnarray*}
C(n+8,0)& = & C(2,0) \otimes C(0,n+6) \\
{} & = & C(2,0) \otimes C(0,2) \otimes C(n+4,4) \\
{} & = & C(2,0) \otimes C(0,2) \otimes C(2,0) \otimes C(0,n+2) \\
{} & = & C(2,0) \otimes C(0,2) \otimes C(2,0) \otimes C(0,2) \otimes 
C(n,0)  \\
{} & = & M(2,\RR) \otimes \HH \otimes M(2,\RR) \otimes \HH \otimes 
C(n,0) \\
{} & = & M(4,\RR) \otimes M(4,\RR) \otimes C(n,0) \\
{} & = & M(16,\RR) \otimes C(n,0)
\end{eqnarray*}
The final results shows that  $C(n+8,0)$ is of the same type as $C(n,0)$; of course, dimensions differ (one has to tensor multiply by real $16 \times 16$ matrices).

In the same way, one obtains (suppose $q < p$) :
\begin{eqnarray*}
C(p,q) & = & C(1,1) \otimes C(p-1,q-1) \\
{} & = & C(1,1) \otimes C(1,1)\otimes \ldots \otimes C(p-q,0) \\
{} & = & M(2^q,\RR)\otimes C(p-q,0)
\end{eqnarray*}
The two results $$C(n+8,0) =  M(16,\RR) \otimes C(n,0) $$ and 
$$C(p,q)=  M(2^q,\RR)\otimes C(p-q,0)$$ show that it is enough to analyze the purely Euclidean case, and that
the classification depends only on $(p-q) \, \hbox{modulo}\,  8$.

\paragraph{Conclusions.}
\begin{center}
\begin{tabular}{|l||c|c|c|c|}
\hline
$p-q \,\hbox {mod}\, 8 $ & $ 0 $ & $ 1 $ & $ 2 $ & $ 3 $\\
\hline
$C(p,q)  $ & $ M(2^{n/2},\RR) $ & $M(2^{{n-1 \over 2}},\RR)\oplus 
M(2^{{n-1 \over 2}},\RR)  $ & $ M(2^{n/2},\RR) $ & $ M(2^{{n-1 \over 
2}},\CC) $ \\
\hline
\hline
$p-q \, \hbox {mod} \, 8$  & $ 4 $ & $ 5 $ & $ 6 $ & $ 7 $ \\
\hline
$C(p,q) $ & $ M(2^{{n\over 2}-1},\HH) $ &$ M(2^{{n-1 \over 2}},\HH) 
\oplus M(2^{{n-1 \over 2}},\HH) $ & $ M(2^{{n\over 2}-1},\HH) $ & 
$M(2^{{n-1 \over 2}},\CC)$ \\
\hline
\end{tabular}
\end{center}

In particular, we have the following tables (we write $(d,K)$ instead of $M(d,K)$) :

{\vskip 0.3cm}
Euclidean case (from $n=4$ to $n=11$). 

{\vskip 0.3cm}
\begin{center}
\begin{tabular}{|c||c|c|c|c|c|}
\hline
$n$ & $C^{{\scriptstyle \CC}}$ & $C(n,0)$ & $C(0,n)$ & $C_0(n,0)$ & 
$\theta$ \\
\hline
$4$ & ${}(4,\CC)$                   &  ${}(2,\HH)$                 & 
${}(2,\HH)$ &  ${}\HH \oplus \HH$ & $ \epsilon $\\
$5$ & ${}(4,\CC) \oplus {}(4,\CC)$  & ${}(2,\HH) \oplus {}(2,\HH)$ &  
${}(4,\CC)$&  ${}(2,\HH)$ & {}  \\
$6$ & ${}(8,\CC)$                   &   ${}(4,\HH)$                & 
${}(8,\RR)$&   ${}(4,\CC)$ & $i \epsilon $\\
$7$ & ${}(8,\CC)\oplus {}(8,\CC)$   &   ${}(8,\CC)$                & 
${}(8,\RR)\oplus {}(8,\RR)$&   ${}(8,\RR)$ & {} \\
$8$ & ${}(16,\CC)$                  &   ${}(16,\RR)$               &  
${}(16,\RR)$&  ${}(8,\RR)\oplus {}(8,\RR)$ & $  \epsilon $ \\
$9$ & ${}(16,\CC)\oplus {}(16,\CC)$ &  ${}(16,\RR)\oplus {}(16,\RR)$ 
& ${}(16,\CC)$&   ${}(16,\RR)$ {} \\
$10$ & ${}(32,\CC)$                 &  ${}(32,\RR)$                & 
${}(16,\HH)$&   ${}(16,\CC)$ & $ i \epsilon$ \\
$11$ & ${}(32,\CC)\oplus {}(32,\CC)$&  ${}(32,\CC)$                & 
${}(16,\HH)\oplus (16,\HH)$&   ${}(16,\HH)$ {} \\
\hline
\end{tabular}
\end{center}
{\vskip 0.3cm}
Hyperbolic case (from $n=4$ to $n=11$). 
{\vskip 0.3cm}
\begin{center}
\begin{tabular}{|c||c|c|c|c|c|}
\hline
$n$ & $C^{{\scriptstyle \CC}}$ & $C(n-1,1)$ & $C(1,n-1)$ & $C_0(n-1,1)$ & 
$\theta$ \\
\hline
$4$ & ${}(4,\CC)$ &  ${}(4,\RR)$ &  ${}(2,\HH)$ &   ${}(2,\CC)$ & $ i 
\epsilon $\\
$5$ & ${}(4,\CC) \oplus {}(4,\CC)$ &  ${}(4,\CC)$ &  ${}(2,\HH) 
\oplus {}(2,\HH)$ &   ${}(2,\HH)$ & {}  \\
$6$ & ${}(8,\CC)$ &  ${}(4,\HH)$ &  ${}(4,\HH)$ &   ${}(2,\HH)\oplus 
{}(2,\HH)$ & $ \epsilon $\\
$7$ & ${}(8,\CC)\oplus {}(8,\CC)$ &  ${}(4,\HH)\oplus {}(4,\HH)$ &  
${}(8,\CC)$ &   ${}(4,\HH)$ & {} \\
$8$ & ${}(16,\CC)$ &  ${}(8,\HH)$ &  ${}(16,\RR)$ &   ${}(8,\CC)$ & $ 
i \epsilon $ \\
$9$ &  ${}(16,\CC)\oplus {}(16,\CC)$ &  ${}(16,\CC)$ &  
${}(16,\RR)\oplus {}(16,\RR)$ &   ${}(16,\RR)$ {} \\
$10$ & ${}(32,\CC)$ &  ${}(32,\RR)$ &  ${}(32,\RR)$ &   ${}(16,\RR) 
\oplus {}(16,\RR)$ & $  \epsilon$ \\
$11$ &  ${}(32,\CC)\oplus {}(32,\CC)$ &  ${}(32,\RR) \oplus 
{}(32,\RR)$ &  ${}(32,\CC)$ &   ${}(32,\RR)$ {} \\
\hline
\end{tabular}
\end{center}
{\vskip 0.3cm}
Looking at these tables, one sees that
 $C(0,n) \neq C(n-1,1) {\hskip 0.2cm} \hbox{ but}{\hskip 0.2cm} 
C(n,0)=C(1,n-1)$
We have also $C_0(n,0)=C_0(0,n)=C(0,n-1)$ and
$C_0(n-1,1)=C_0(1,n-1)=C(1,n-2)$

These are particular instances of the general relation
 $$C_0(p,q) = 
C(p,q-1)=C(q,p-1)=C_0(q,p)$$

\paragraph{The structure of complex Clifford algebras.} $\quad$

One defines the chirality operator $\theta$ (often called ``$\gamma^5$'')  as follows.
If $\epsilon^2=1$, one sets $\theta = \epsilon$ whereas If $\epsilon^2=-1$, one sets $\theta = i \epsilon$.
In all cases  $\theta^2 =1$. Therefore, in all cases (using complex numbers), one can build the two projectors
$P_{L,R} = \frac{1 \pm \theta}{2}$.

Now the periodicity is only modulo $2$. The proof is immediate. Set $f=2^{[n/2]}$.
Then,

\begin{eqnarray*}
&\hbox{ n even} & \quad C^{\CC} \simeq M(f,\CC) \,  \hbox{,  it is simple } \\
&{}& \quad C_0^{{ \CC}} = {1 - \gamma_5 \over 2}C_0^{{ 
\CC}} \oplus  {1 + \gamma_5 \over 2}\,  C_0^{{ \CC}} \,  \hbox{, it is not  simple} \\
&\hbox{ n odd} & \quad  C_0^{{ \CC}}  \, \hbox{, it  is simple} \\
&{}& \quad C^{{ \CC}} = {1 - \gamma_5 \over 2}C^{{ \CC}} 
\oplus  {1 + \gamma_5 \over 2}\,  C^{{\CC}} \simeq M(f,\CC) \oplus M(f,\CC) \,  \hbox{, it is not simple} 
\end{eqnarray*}

\paragraph{Terminology.}

\begin{description}
\item[{\sl Dirac spinors}]
This is an irreducible representation of the complexified Clifford algebra $C^{{\scriptstyle 
\CC}}=C(p,q)^{{\scriptstyle \CC}}$. It is faithful (and unique) when $n=p+q$ is even;  it is not faithful when $n$ is odd.
In all cases $E_{Dirac} = \CC^f$ where $f=2^{[n/2]}$.

\item[{\sl Weyl spinors}]
This is defined only when $n$ is even, as an irreducible representation of the complexified  Clifford even subalgebra. When $n$ is even, we have indeed the reduction 
$E_{Dirac} =  E_{Weyl}^L \oplus  E_{Weyl}^R$. Left and right Weyl spinors are defined by

$$ E_{Weyl}^L = \{\psi \in E_{Dirac} \, \vert\, ({1 - \gamma_5 \over 
2}\psi=\psi
\Leftrightarrow \gamma_5 \psi = -\psi \}$$
$$ E_{Weyl}^R = \{\psi \in E_{Dirac} \, \vert \, ({1 + \gamma_5 \over 
2}\psi=\psi
\Leftrightarrow \gamma_5 \psi = \psi \}$$
One can always decompose $\psi \in E_{Dirac} : \psi = 
\psi_L + 
\psi_R$ with $\psi_L = ({1 - \gamma_5 \over 2})\psi$ and $\psi_R =
({1 + \gamma_5 \over 2})\psi$.

When $n$ is odd, the restriction of the Dirac representation from $C^{{\scriptstyle \CC}}$ to $C_0^{{\scriptstyle \CC}}$ stays irreducible.

We remind (we shall come back to this later) the reader that  groups $Pin(p,q)$ , $Spin(p,q)$ and $Spin_0(p,q)$ 
can be realized  as subsets of the algebras$C(p,q)$ or $C_0(p,q)$. 
The Dirac and Weyl spinors also provide representations for these groups.

\item[ {\sl Majorana spinors}]

Given a dimension $n$ and a pair of integers $(p,q)$ with
$n=p+q$, people say that 
 {\sl  Majorana spinors\/} exist whenever {\it one of the two\/} algebras
$C(p,q)$ or $C(q,p)$ is of real type.

Existence of Majorana spinors can be read from the previous tables.

In all such cases, one can find an anti-linear operator $c$ of square $1$ such that 
$$E_{Majorana} = \{\psi \in E_{Dirac}\, \vert \quad c\psi = \psi \}$$

\item[{\sl Weyl-Majorana spinors (we suppose $n$ even)}]

Notice that $c$ (when it exists)  and $\theta $  commute if $\theta = \epsilon$,   whereas
 $c$  and $\theta $ anti - commute if $\theta = i \epsilon$. When they commute, the two conditions can be imposed simultaneously. This happens only when 
 $(p-q) = 0 \,\hbox{mod} \, 8$, \ie  
if the space is Euclidean,  in dimensions $8, 16, 24\ldots$ and, 
If the space is Lorentzian,  in dimensions $2, 10, 18 \ldots$ 
In those cases, 
one defines  {\sl Weyl-Majorana spinors\/} of two possible chiralities (left or right), as follows :
$$ E_{Majorana}^L = \{\psi \in E_{Weyl}^L   \vert \quad c\, 
\psi=\psi\}$$
$$ E_{Majorana}^R = \{\psi \in E_{Weyl}^R \vert \quad c \, 
\psi=\psi\}$$

\item[Remarks] $\quad$

In particle physics, Dirac spinors describe the ``most''  elementary  {\sl charged} particles (like the electron).
The charge conjugation operator associates each particle described by the spinor $\psi$ with its antiparticle described by the spinor $c(\psi)$. Explicitly (\ie using  components, after having selected a basis in the complex vector space)  the antilinear operator $c$ appears as the product of the complex conjugation operator with a linear operator $C$ whose expression depends on the explicitly chosen  representation (calculations can be performed  intrinsically by using the $c$ operator,  and this is much simpler than using matrices !)

In the framework of the standard model of particle physics, 
neutrinos are supposed to be massless and are described by left handed Weyl spinors. However, it seems nowadays that neutrinos have a mass, and in that case they are described by Dirac spinors...but the two Weyl components of the neutrinos have anyway very distinct types of interactions :  in particular the right handed component is only coupled to the Higgs fields, not to the gauge bosons. Incidentally this shows that ``nature''  is not parity invariant. 

Although Majorana spinors are mathematically available in dimension $(3,1)$, they do not enter in the formulation of the standard model... and, in this sense,  do not seem to correspond to  known elementary particles.

There are no   Weyl-Majorana spinors in dimension $4$ when the signature is Euclidean or Lorentzian (they would mathematically exist for a signature $(2,2)$).

Algebras $C(3,1)$ and $C(1,3)$ are not isomorphic. The first is faithfully realized on $\RR^4$ and the next on the vector space $\HH^2$ of bi-quaternions. These representations do not seem to play any role in four dimensional space-time physics. The above two Clifford algebras share the same even subalgebra $C_0(3,1)  =  C_0(1,3)$, of course.
 \end{description}

\section{Spin groups}
When the signature is (purely) Euclidean, the group $L=O(n)$ has two connected components (distinguished by the sign of the determinant):  $L = L_+  \cup L_-$. When $n=(p,q)$ refers to an arbitrary signature, $L = O(p,q)$ has four connected components ( $L = L_+^\uparrow \cup L_+^\downarrow \cup L_-^\uparrow \cup 
L_-^\downarrow $) since one has to take into account  the existence of space - orientation {\sl and \/} time - orientation preserving, or non - preserving, isometries.

In the Euclidean case, the $SO(n)$ group is connected but not simply connected and the group $Spin(n)$ is its universal double cover\footnote{There are several well-known  isomorphisms in low dimension, we mention;
$Spin(3)=SU(2)$, 
$Spin(4)=SU(2)\times SU(2)$, 
$Spin(5)=U(2,\HH) = 
USp(4)$, $Spin(6)=SU(4)$,
 $Spin^{\uparrow}(2,1)=SL(2,\RR)$,
 $ Spin^{\uparrow}(3,1)=SL(2,\CC)$,
$Spin^{\uparrow}(4,1)=U(1,1,\HH)$,
$Spin^{\uparrow}(5,1)=SL(2,\HH), Spin^{\uparrow}(3,2)=Sp(4,\RR), 
Spin^{\uparrow}(4,2) = SU(2,2).$}: $Spin(n)$ is a principal fiber bundle, with structure group $Z_2$  over $SO(n)$. Otherwise, for signatures $n=(p,q)$, the group $SO(n)$ itself is not connected, and one should consider the identity component $L_+^\uparrow = SO^\uparrow(p,q)$ (isometries that preserve both space {\sl and} time orientation) and its double cover  $Spin^\uparrow(p,q)$. 

We now show how one can explicitly realize the Spin group in the Clifford algebra.
First we embed the vector space $E$, with signature $(p,q)$, in the algebra $C(p,q)$, by associating the vector $v=v_\mu e^\mu$ (where $\{e^\mu\}$ is a basis) with the element $v = v_\mu \gamma^\mu$ of the Clifford algebra (this element is often denoted $\slashed{v}$, after R. Feynman).

{\bf The Clifford group} $\quad$
The Clifford group $\Gamma$ is defined as the set of all elements $s$ of $C=C(p,q)$ that are invertible and are such that 
$$ \forall x \in E, \quad s x s^{-1} \in E $$

Every invertible element of $E$ belongs to the Clifford group. Indeed : 
$$
vxv^{-1} = {1\over v^2}vxv = {1\over v^2}(-vvx + 2 g(v,x))=-x+2g(v,x) 
v /v^2
$$

This is the opposite of a symmetry with respect  to a hyperplane.

More generally, for  $s \in \Gamma$,  the transformation $\chi(s): x \in E \mapsto s x s^{-1}$ belongs to the orthogonal group $O(p,q)$. The proof stems from the fact that every rotation can be obtained as a product of symmetries with respect to hyperplanes.

{\bf The Pin and Spin groups} $\quad$

First we define the Dirac ``bar'' operation on the generators (take also complex conjugate if there are complex coefficients) :
$${\overline {\gamma^1 \gamma^2 \ldots  \gamma^p}}= \gamma^p  \ldots  
\gamma^2 
 \gamma^1$$

The Clifford group is  slightly ``too big'', so,  one introduces a normalization condition\footnote{The name was given by  J.P. Serre (untranslatable french joke)}

$$
Pin= \{ s \in  \Gamma / \quad  \vert \overline{s} s \vert =  1\}
$$
Elements of $Pin$ can be written as products $s=u_1 u_2 \ldots u_k$ where the $u_i$ are (co)-vectors of the vector space $E$, embedded in its Clifford algebra as described above, and such that 
\begin{eqnarray*}
k \, \hbox {even and}  \quad {\overline {s}s} > 0 &\Longleftrightarrow & 
\chi(s) \in L_+^\uparrow \\
k \, \hbox {even andt}  \quad {\overline {s}s} < 0 &\Longleftrightarrow & 
\chi(s) \in L_+^\downarrow \\
k \, \hbox {odd and}  \quad {\overline {s}s} > 0 &\Longleftrightarrow & 
\chi(s) \in L_-^\uparrow \\
k \, \hbox {odd and}  \quad {\overline {s}s} < 0 &\Longleftrightarrow & 
\chi(s) \in L_-^\downarrow 
\end{eqnarray*}
so that
$$
Spin^\uparrow = \{ s \in  \Gamma \cap C_0, / \overline{s} s = 1\}
$$

 Notice that $Pin(p,q)$ is not necessarily isomorphic with $Pin(q,p)$ but the equality holds for $Spin$.

 If $s \in Spin$, then $-s \in Spin$ as well, and $\chi(s)=\chi(-s)$, so that $Ker \, \chi = \{-1,+1\}= \ZZ_2$, as it should.

{\bf The Lie algebra of the Spin group} $\quad$

Take $x,y \in E$.

\begin{eqnarray*}
& exp(xy) \in Spin^\uparrow & \Leftrightarrow exp({\overline {xy}})exp(xy) = 1 \\
&{}& \Leftrightarrow
 exp(yx+xy) = 1  \Leftrightarrow xy + yx = 0 \\
&{}& \Leftrightarrow g(x,y) = 0 
\end{eqnarray*}
So, $z \in C(p,q)$ belongs to $Lie(Spin^\uparrow)$
if and only if it can be written as a linear combination of products $xy$ where
$x$ and $y$ are orthogonal, \ie by the products $\gamma^\mu \gamma^\nu$.
We recover the fact that
$dim(Lie(Spin(p,q)))_{\vert{n=p+q}}=n(n-1)/2$.

Example. Take $(p=3,q=1)$. Then for instance ${\beta \over 2} \gamma^0 \gamma^1 \in 
Lie(Spin^\uparrow)$ and  $s \doteq exp({\beta \over 2} \gamma^0 
\gamma^1) \in Spin^\uparrow$.
An easy calculation (expand the exponential term) shows that
$$s = \cosh {\beta \over 2} + \gamma^0 \gamma^1 \sinh  {\beta \over 
2}$$
This describes a Lorentz transformation (boost) along the first axis. Indeed, 
\begin{eqnarray*}
&s\gamma^0s^{-1} &= \gamma^0 \cosh \beta + \gamma^1 \sinh \beta \\
&s\gamma^1s^{-1} &=  \gamma^0 \sinh \beta + \gamma^1 \cosh \beta \\
&s\gamma^is^{-1} &= \gamma^i \, \hbox{pour} \,  i = 2,3
\end{eqnarray*}
Remember that, in special relativity, $th(\beta) = v/c$, where $c$ is the speed of light and $v$ is the speed that defines the boost.

More generally, if
$$
\mbox{$
s=exp({\beta \over 2} \gamma^0 ({\vr v}.{\vr \gamma}))
$}
$$ $\chi(s)$ 
is a boost of parameter $\beta$ along the direction 
$\overrightarrow v$
$$
\mbox{$
s= exp({\theta \over 2}  {\vr n}.{\vr \gamma})
$}
$$ $\chi(s)$ 
is a rotation of angle $\theta$ around the direction ${\vr n}$, with 
($\vert n \vert^2 = 1$).

\section{Spin bundles and spin connection}

\subsection{Bundle of spinors and spinor fields}

A spinorial structure (when it exists) is an extension of  the principal bundle of orthonormal frames for the (pseudo) riemannian manifold $(M,g)$. Since $SO(n)$ is a quotient of $Spin(n)$, existence of the extension is not automatic (change of structure group from a group $H$ to a group $G$ is automatic when $H$ is a subgroup of $G$, not when $H$ is a quotient of $G$). In the present case,  the obstruction is measured by the so-called second Stiefel-Whitney  characteristic class, but from now on we suppose that this potential difficulty does  not appear. 

The metric being chosen, call $P = P(M,SO(n))$  its bundle of orthonormal frames.
A spinorial structure $\hat P$ is therefore a principal bundle $\hat P = \hat P(M,Spin(n))$, with a bundle homomorphism $\chi$ from 
$\hat P$ to $P$. This means that $\chi$ preserves the fibers and commutes with the action of the corresponding structure groups : calling again  $\chi : Spin(n) \mapsto SO(n)$ the usual double covering map, one imposes the following : if  $\hat z \in \hat P$ and $\hat k \in Spin(n)$, then we have $\chi(\hat z \hat k) = \chi(\hat z) \chi(\hat k)$. 

It is often convenient to use the Spin bundle $\hat P$ to construct a bundle of Clifford algebras over the manifold $M$.

$\hat P$ is a principal bundle, so that if we consider a representation of its structure  group $Spin (n)$ on the vector space $\CC^s$, we can build the associated vector bundle $SM = \hat P \times_{Spin(n)} \CC^s$.
Any representation of $Spin(n)$ can be used, but it is standard to use either Dirac spinors (this is a reducible representation of the spin group in even dimensions), or one of the two Weyl representations. The bundle $SM$ is respectively called the bundle of Dirac spinors or the bundle of Weyl ( left or right) spinors. Sometimes Weyl spinors are called ``half-spinors''.

The sections of the vector bundle $SM$ are called the spinor  fields.

\subsection{Spin connections}

Given a (pseudo) riemannian metric $g$ on the manifold $M$, a {\sl metric connection\/} is a principal connection on the bundle of orthonormal frames $P = P(M,SO(n))$. It may have torsion.  If it does not, it is called the Levi-Civita connection (or ``the'' riemannian connection). Choose a metric connection and call $\Gamma = \Gamma_\mu dx^\mu$  the connection one-form on the manifold $M$,  the base of the bundle $P$ (a local section is therefore chosen); 
it is valued in the Lie algebra of the group $SO(n)$ : $\Gamma_\mu = \Gamma^a_{b \mu} \, X_a^b$, where 
$X_a^b$ are the generators of this group (antisymmetric matrices).  The connection matrix (a matrix of one-forms)  is $\Gamma^a_b =  \Gamma^a_{b \mu} \, dx^\mu$. One also sets $\Gamma_{ab} = g_{a a^\prime} \Gamma^{a^\prime}_b$.

 Indices $\{a, b, \ldots\}$ can be identified with those of a local orthonormal moving frame $\{e_a\}$ on $M$, with $e_a = e_a^\mu \partial_\mu$  (often called ``vielbein'' by physicists), so that the connection is also characterized by the symbols $\Gamma^a_{b c} = e_c^\mu \Gamma^a_{b \mu}$. 
 
A spin connection is a principal connection on the bundle $\hat P$. There is bijection between the set of connections on $P$ and the set of connections on $\hat P$. If $\chi$ denotes the fiber homomorphism from $\hat P$ tp $P$, and if $\hat \omega$ is the connection that covers $\omega$, we have
$\hat \omega (\overrightarrow z) =  \omega (\chi_*(\overrightarrow z) )$ where $\overrightarrow z$ is a tangent vector and $\chi_*$ is the tangent map of $\chi$.  Call $M=M^a_b X_a^b$ an arbitrary matrix of $Lie(SO(n))$; $M^a_b$ are coefficients and $X_a^b$ are generators.  Of course $n$ may refer to a metric $g$ of arbitrary signature $(p,q)$ and we have $M=M^a_b X_a^b = M_{ab}X^{ab}$ with $M_{ab} = g_{aa^\prime} M^{a^\prime}_b$ and $X^{ab}= g^{aa^\prime}X_{a^\prime}^b$.  Define
$\chi(M) = \chi(M_{ab}X^{ab}) = M_{ab} \,  \chi(X^{ab}) = M_{ab} \,   \frac{1}{8} [\gamma^a, \gamma^b] = M_{ab} \,   \frac{1}{4} \gamma^a \gamma^b$, that belongs to $Lie(Spin(n))$. With this definition, 
one can check the Lie algebra homomorphism property $[\chi(M_1) , \chi(M_2) ] = \chi([M_1 M_2])$.

The spin connection  therefore reads $\hat \Gamma = \frac{1}{4} \Gamma_{ab} \gamma^a \gamma^b$.
The corresponding covariant differential acting on spinor fields (its values are spinorial one-forms) is
$ \nabla \Psi = d \Psi +  \frac{1}{4} \Gamma_{ab} \gamma^a \gamma^b \Psi$.
The Dirac operator is ${\slashed{\nabla}} \Psi = \gamma^a \nabla_a \Psi$ where  $\nabla_a \Psi = < \nabla \Psi, e_a> $ is the covariant derivative of $\Psi$ in the direction $e_a$. 

In Physics, spinors are not only coupled to gravity (described by the spin connection) but also to electromagnetic, weak or strong forces that are themselves described by appropriate Lie algebra valued connection one-forms leading to generalized expressions for the Dirac operator. 

\section{Spinors and fundamental interactions}

We now summarize how  spinors are used in the standard model of elementary particles  ($3+1$ dimensional physics). Space-Time is described as a four-dimensional manifold endowed with a metric of signature $(3,1)$ or $(1,3)$. It is  both space and time orientable and oriented (hence totally oriented). The isometry group $SO^\uparrow (3,1)$ is called  Lorentz group,  its spin covering is isomorphic with $SL(2,\CC)$.  The  Clifford algebras  $Cliff^{\CC} = M(4,\CC)$ 
(also called ``Dirac algebra'')  is the common complex extension of the relevant two real forms $Cliff(3,1)=M(4,\RR)$ and $Cliff(1,3) = M(2, {\HH})$. The algebra $Cliff^{\CC}$ is simple; its even part (not simple) is isomorphic with $M(2,\CC) \oplus M(2,\CC)$. The Dirac spinor fields $\Psi$ are valued in $\CC^4$. The Weyl spinors (``half-spinors''   $\Psi_L$  or  $\Psi_R$ ) fields take values in $\CC^2$. Dirac spinors and Weyl spinors are used to describe the most fundamental particles (those that appear in the Standard Model). Majorana spinors and bi-quaternions, although mathematically available in this dimension, are not used in the Standard Model to describe elementary particles. The orientation operator $\epsilon = \gamma^0 \gamma^1 \gamma^2 \gamma^3$ is of square $-1$. The chirality operator $\theta$ in this dimension is usually called $\gamma_5$ and is defined by $\gamma_5 = i \epsilon$;  the complex $i$ is necessary to make its square equal to $+1$. For this reason, $\gamma_5$  does not commute with charge conjugation. 

The fundamental elementary particles :
\begin{itemize}
\item 6 leptons in three families : the electron and its neutrino, the muon and its neutrino, the tau and its neutrino. 
\item 6 quarks in three families : the up and down quarks, the charm and strange quarks,  the top and bottom quarks (also called truth and beauty).
\item The gauge bosons : the photon, the weak bosons $W_+, W_-,  Z$ and the eight gluons.
\item The Higgs boson (a so called ``scalar particle'')
\item The graviton
\end{itemize}

As a rule, matter fields are described by sections of associated bundles over the space-time manifold. 
The structure group of the principal bundle is $Spin(3,1) \times SU(3) \times SU(2) \times U(1)$.
The group $SU(3)$ describes the color interactions (``chromodynamics''), \ie  strong interactions, and $SU(2)\times U(1)$ describes  electroweak interactions.

Leptons, as quarks, are associated with the fundamental representations of $Spin(3,1)$, so they are described by spinors. From the space-time point of view, there is no difference between leptons and quarks : the difference lies in the coupling to the other structure groups (the covariant derivatives -- involving the connection forms --  relative to  appropriate representations of $SU(3) \times SU(2) \times U(1)$).

The other ``elementary particles'' are bound states of fundamental elementary particles (sections of higher tensorial bundles).  In general they are bound states of quarks (like the proton, neutron,  \ldots pions, kaons \ldots \etc.).

If we forget  strong and electroweak interactions,  the three  leptons (electron, muon and tau) and the six quarks are described by Dirac spinors (four complex components when some spinorial moving frame has been chosen). Each such Dirac spinor can be written as the sum of the two Weyl spinors.
The three neutrinos used to be described by left Weyl spinors, but it seems nowadays experimentally clear that they are massive, so they are now described, like the others leptons, by Dirac spinors. 

The $SU(3)$ and $SU(2) \times U(1)$ gauge bosons are Lie algebra valued one-forms on space-time and describe the different connections. Their writing as such (on the base rather than on the total space)  involves the choice of  local sections in the corresponding principal bundles (``gauge choice''). 

Scalar particles are, by definition, associated with the trivial representation of the Lorentz group.  Using ``traditional'' differential geometry the Higgs boson has therefore to be described as a scalar field (non trivially coupled to the $SU(2)\times U(1)$ structure group). However, in the framework of non-commutative geometry  -- that will not be discussed here --  the Higgs field can also be considered as a component of some generalized connection form.

The graviton is described as the metric tensor field itself.

Couplings (``forces'') between particles stem from the covariant derivatives acting on sections of vector bundles (matter fields). These couplings appear in the Dirac equations  obeyed by those fields.

Yang-Mills equations (generalizing Maxwell equations), Einstein equations (describing gravity),  and the various types of Dirac equations describing the couplings of elementary particles,  can be derived from a Lagrangian density.
This (real) scalar quantity has to be Lorentz invariant. In particular, the term $(\Psi, \slashed{\nabla} \Psi)$ 
 describing the Dirac equation should involve a spin-invariant scalar product in the space of Dirac spinors.  
 This scalar product (called the Dirac scalar product) differs from the naive hermitian scalar product in the representation space. Explicitly, if $\Psi$ and $\Phi$  denote  spinors with four explicit complex components, the hermitian scalar product reads  $\Psi^\dag \Phi$ whereas the Spin invariant scalar product reads  $\overline \Psi \Phi$ 
 where $\overline \Psi = \Psi^\dag \gamma^0$. More generally, \ie for arbitrary signatures, one has to plug in the product (in some order) of the $\gamma$ generators associated with the time-like directions\footnote{The free Dirac lagrangians, for signatures $(p,q)$ and $(q,p)$,  differ by $i$ and sign factors}. Spin invariant scalar products may  have bigger isometry group than the group Spin itself; for a usual signature of space-time,  the Dirac scalar product is hermitian symmetric,  neutral,  and has invariance group $U(2,2,\CC)$. Of interest is also the conformal group, whose double cover, in dimension $(p,q)$ is isomorphic with $Spin(p+1, q+1)$, \ie in the case of Space-Time,  with $Spin(4,2)$.

Terminology: 
In  the framework of the electroweak and strong interactions, 
the words ``singlet'' and  ``doublet'' refer respectively to the trivial representation and to the fundamental representation of the group $SU(2)$, 
the word ``weak hypercharge'' refers to the number $Y$  characterizing the representation chosen for the $U(1)$ component of $SU(2) \times U(1)$, 
the word ``color triplet'' refers to the fundamental representation of the group $SU(3)$.
 With this terminology, we have the following description for the different couplings of fundamental elementary particles.

One introduces a doublet of left-handed Weyl spinors for each family of leptons and quarks: 

$$\begin{pmatrix}  u_L \\  d_L\end{pmatrix} , 
\begin{pmatrix}  c_L \\  s_L\end{pmatrix} , 
\begin{pmatrix}  t_L \\  b_L\end{pmatrix} , 
\qquad
\begin{pmatrix}  e_L \\  \nu_L^e\end{pmatrix} , 
\begin{pmatrix}  \mu_L \\  \nu_L^\mu \end{pmatrix} , 
\begin{pmatrix}  \tau_L \\  \nu_L^\tau\end{pmatrix} $$

The corresponding weak hypercharges are $1/3, 1/3, 1/3$ for the quarks and $-1, -1, -1$ for the leptons.

One also introduces a right-handed Weyl spinor (a singlet) for each  lepton (we have six of them) and for each quark (also six of them) : 

$$ u_R, d_R, c_R, s_R, t_R, b_R \qquad  e_R, \nu_R^e, \mu_R, \nu_R^\mu, \tau_R, \nu_R^\tau$$

The corresponding weak hypercharges are respectively $4/3, -2/3, 4/3, -2/3, 4/3, -2/3$ for the quarks and $-2, 0, -2, 0, -2, 0$ for the leptons.

For  each of the three families, and both for quarks and leptons,  the sum of weak hypercharges for left handed particules is equal to the corresponding sum for right handed particles. Notice also that left and right components (Weyl spinors) of the  three neutrinos have very different couplings to the $SU(2) \times U(1)$ connection:   their  right components are $SU(2)$ singlets and have a weak hypercharge equal to  $0$: they are not coupled at all to the electroweak gauge fields. 

To complete the picture, one has to triplicate each of  the quarks,  introducing a so-called color triplet coupled\footnote{For instance, a left-handed down quark of given color  is an element of the tensor product  $\CC^2 \otimes \CC^3 \otimes \CC^2 \otimes \CC $, which is a representation space for $Spin(3,1) \times  SU(3) \times SU(2)  \times U(1) $.}  to the $SU(3)$ connection (one triplet of Dirac spinors, or two triplets of Weyl spinors).

For a space-time signature, if $\Psi$ is a Dirac spinor and $A$ is a one-form, one observes that $\overline \Psi \gamma^\mu A_\mu \Psi = \overline \Psi_L \gamma^\mu A_\mu \Psi_L + \overline \Psi_R \gamma^\mu A_\mu \Psi_R$, which means that the connection does not couple the  left and right Weyl spinors : in absence of scalar interactions, the left-handed and right-handed worlds ``do not talk'' to each other. However, if $\phi$ is a scalar field (like the Higgs field), we see, at the contrary, that  $\overline \Psi \phi \Psi = \overline \Psi_L \phi \Psi_R + \overline \Psi_R \phi \Psi_L$. 
 This observation is basically at the root of the mechanism that gives a mass to particles : if the scalar field $\phi$ is expanded around some non-zero numerical value, we obtain a mass term for a Dirac spinor, ie a term proportional to 
 $\overline \Psi  \Psi$. 
 
There are formulations of the Standard Model where scalar fields are components of a generalized connection and where the two kinds of interactions can therefore be unified :  for instance,  one can use particular superconnections associated with the Lie supergroup  $SU(2|1)$ (here the word ``super'' refers  to ``left-right'' and not at all to ``boson-fermion''); one can also use a formalism based on non-commutative geometry. These are very nice reformulations of the Standard Model (suggesting possible new developments) but they describe the same Standard Model.

\section{Comments}

The subject of Clifford algebras and spinors is not new... It seems that, at some point,  it was even almost forgotten by many mathematicians. However,  during the years 1940-1970,   it became such a fundamental tool of particle physics that every PhD student in theoretical physics had to be able of performing rather involved calculations in the Dirac algebra, or more generally, in  ``Clifford algebras  bundles'' (without necessarily being aware of the mathematical terminology). Several powerful languages of computer algebra,  for instance Reduce or Stensor, both written in LISP, or Schoonship, written in assembler language, had dedicated packages for performing such algebraic manipulations  -- and all this was already available at the end of the sixties.  Such programs have been used  by generations of high energy physicists doing calculations in quantum field theory (in particular in quantum electrodynamics\footnote{Almost any calculation in quantum electrodynamics, for instance, involves taking products and traces in a Clifford algebra bundle and performing $4$-dimensional iterated integrals.}).
The subject of spinors and Clifford algebra came back, later,  at the forefront of differential geometry and of mathematics in general, with the recognition of the importance of the Dirac operator in  geometry, and the discovery of the Atiyah - Singer theorem. All of this now belongs to the standard toolbox of the mathematician.  Spinors may not have yet revealed all their mysteries and they will certainly show up again, one of these days, maybe in a new guise, in Physics and in Mathematics.

 I shall conclude this talk by mentioning several related problems  that are, in my opinion (shared by a number of colleagues !) among the most important questions of contemporary physics : 
 \begin{itemize}
 \item An explanation for the values of the fundamental constants of nature should be found (in particular the masses of elementary particles, the mixing angles, or the  values of the coupling constants).
 \item The necessity of using other types of interactions, like supersymmetries, strings, branes etc,  in our description of physics, is still an unsettled possibility.
 \item Another deep problem is to find a quantum theory of gravity.  Such new developments will certainly involve the concepts of spinors in non commutative geometry... but in which way ?
 \end{itemize}

  \smallskip
  
There are so many references on Clifford algebras that it is hard to give a list.   I shall therefore only mention my own papers  \cite{Coquereaux : periodicityClifford},  \cite{Coquereaux: TexasLecturesClifford},  \cite{Coquereaux-Jadczyk: book}, and quote several articles that can be qualified as   ``historical references'' : \cite{Cartan},  \cite{Chevalley},    \cite{Karoubi}, \cite{Bott}.
  


\begin{thebibliography}{99} 

 \bibitem{Cartan} E. Cartan {\em Le\c cons sur la th\'eorie des spineurs},  Hermann,  Paris, 1938.
 
  \bibitem{Chevalley} C. Chevalley, {\em The algebraic theory of spinors}, Columbia, New York, 1954.
  
      \bibitem{Karoubi}  M. Karoubi, {\em Alg\`ebres de Clifford et K-th\' eorie}, Doctoral thesis, Paris, 1967.
  
   \bibitem{Bott} M.F. Atiyah, R. Bott, A. Shapiro, {\em Clifford Modules } Topology, Vol 3, Sup. 1, 1969.
   
   \bibitem{Coquereaux : periodicityClifford} R. Coquereaux, 
 		{\em Modulo 8 periodicity of Clifford algebras and particle physics},  Phys. Lett. 115 B, 389, 1982.
   
   \bibitem{Coquereaux: TexasLecturesClifford}  R. Coquereaux, {\em Spinors, reflections and Clifford algebras : A review.}, Austin prep. UTTG-21-85,  pub. in ``Spinors in Physics and Geometry'',  W. S., 
 		Eds.  A.Trautman, G. Furlan, 1986.
   
   \bibitem{Coquereaux-Jadczyk: book}
R. Coquereaux and  A. Jadczyk, 
{\em Riemannian geometry, fiber bundles, Kaluza-Klein theories and all
that"}, World Scientific Lecture Notes in Physics, Vol 16. (345 p), 1988.

\end{thebibliography}
\end{document}